\documentclass[aps,twocolumn,preprintnumbers,amsmath,amssymb]{revtex4}

\usepackage{graphicx}
\usepackage{dcolumn}
\usepackage{bm}
\usepackage{multirow}
\usepackage{color}
\usepackage{ulem}
\usepackage{amsmath}
\usepackage{gensymb}

\begin{document}
\title{Exciton diffusion in WSe$_2$ monolayers embedded in a van der Waals heterostructure}

\author{F. Cadiz$^{1}$}
\email{fabian.cadiz@polytechnique.edu}
\author{C. Robert $^{2}$}
\author{E. Courtade $^{2}$}
\author{M. Manca $^{2}$}
\author{L. Martinelli $^{1}$}
\author{T. Taniguchi$^3$}
\author{K. Watanabe$^3$}
\author{T. Amand $^{2}$}
\author{A. C. H.  Rowe $^{1}$}
\author{D. Paget $^{1}$}
\author{B. Urbaszek $^{2}$}
\author{X. Marie $^{2}$}

\affiliation{$^1$ Physique de la mati\` ere condens\' ee, Ecole Polytechnique, CNRS, Universit\' e  Paris Saclay,91128 Palaiseau, France}

\affiliation{$^2$ Universit\'e de Toulouse, INSA-CNRS-UPS, LPCNO, 135 Av. Rangueil, 31077 Toulouse, France}

\affiliation{$^3$National Institute for Materials Science, Tsukuba, Ibaraki 305-0044, Japan}


\begin{abstract}
 We have combined spatially-resolved steady-state micro-photoluminescence ($\mu$PL) with time-resolved photoluminescence (TRPL) to investigate the exciton diffusion in a WSe$_2$ monolayer encapsulated with hexagonal boron nitride (hBN). At 300 K, we extract an exciton diffusion length $L_X= 0.36\pm 0.02 \; \mu$m and an exciton diffusion coefficient of $D_X=14.5 \pm 2\;\mbox{cm}^2$/s. This represents a nearly 10-fold increase in the effective mobility of excitons with respect to several previously reported values on nonencapsulated samples.  At cryogenic temperatures, the high optical quality of these samples has allowed us to discriminate the diffusion of the different exciton species :  bright and dark neutral excitons, as well as charged excitons.  The longer lifetime of dark neutral excitons yields a larger diffusion length of $L_{X^D}=1.5\pm 0.02 \;\mu$m.
\end{abstract}


\maketitle
\textit{Introduction.---}
Two-dimensional crystals of transition metal dichalcogenides (TMDC) such as MX$_2$ (M=Mo, W; X=S, Se, Te) are promising atomically flat semiconductors for applications in nanoelectronics and optoelectronics \cite{Butler:2013a,Geim:2013a,Mak:2010a, Splendiani:2010a, Wang:2012c}. For example, solar cells \cite{Tsai:2014a},photodetectors \cite{Lopez:2013a} and laser  prototypes \cite{Salehzadeh:2015a} based on mono or few-layer MoS$_2$ have been recently  demonstrated. In addition to their potential for unconventional, atomically thin and flexible optoelectronics, the interplay between inversion symmetry breaking and strong spin-orbit coupling in monolayers (MLs) also yields unique spin/valley properties which are expected to provide additional functionalities in future devices \cite{Xiao:2012a,Sallen:2012a,Mak:2012a,Kioseoglou:2012a,Cao:2012a,Jones:2013a,Yang:2015a}.\\

\noindent
 Developments in optoelectronics  based on TMDC MLs need materials with well defined optical transitions and known transport parameters. Due to enhanced Coulomb interaction and large carrier effective masses in these 2D materials, excitons are characterized by large binding energies and are stable even at room temperature \cite{He:2014a,Ugeda:2014a,Chernikov:2014a,Ye:2014a,Qiu:2013a,Ramasubramaniam:2012a,Wang:2015b}. Thus, knowledge on the motion and persistence of inhomogeneous excitonic distributions is thus central to the technologies based on the excitonic properties of TMDC MLs. The relevant parameters are the averaged exciton lifetime $\tau_X$ and diffusion coefficient $D_X$, which determine the time and length scale available for transport and manipulation of a non-equilibrium exciton distribution. \\
 
\noindent
 So far, very few reports exist on the transport dynamics of excitons in TMDC MLs and there is a lack of experimental investigations at cryogenic temperatures where diffusion of different excitonic complexes can be resolved if the optical transitions are narrow enough. In this work, we measure the transport parameters of excitons in ML WSe$_2$ at room and cryogenic temperatures by combining spatially-resolved steady state and time-resolved photoluminescence (PL). At low temperatures, the excellent optical quality of encapsulated MLs \cite{Cadiz:2017a} has allowed us to resolve the diffusion of both the bright and dark neutral excitons, as well as the charged excitons for the first time.

\begin{figure}
\includegraphics[width=0.5\textwidth]{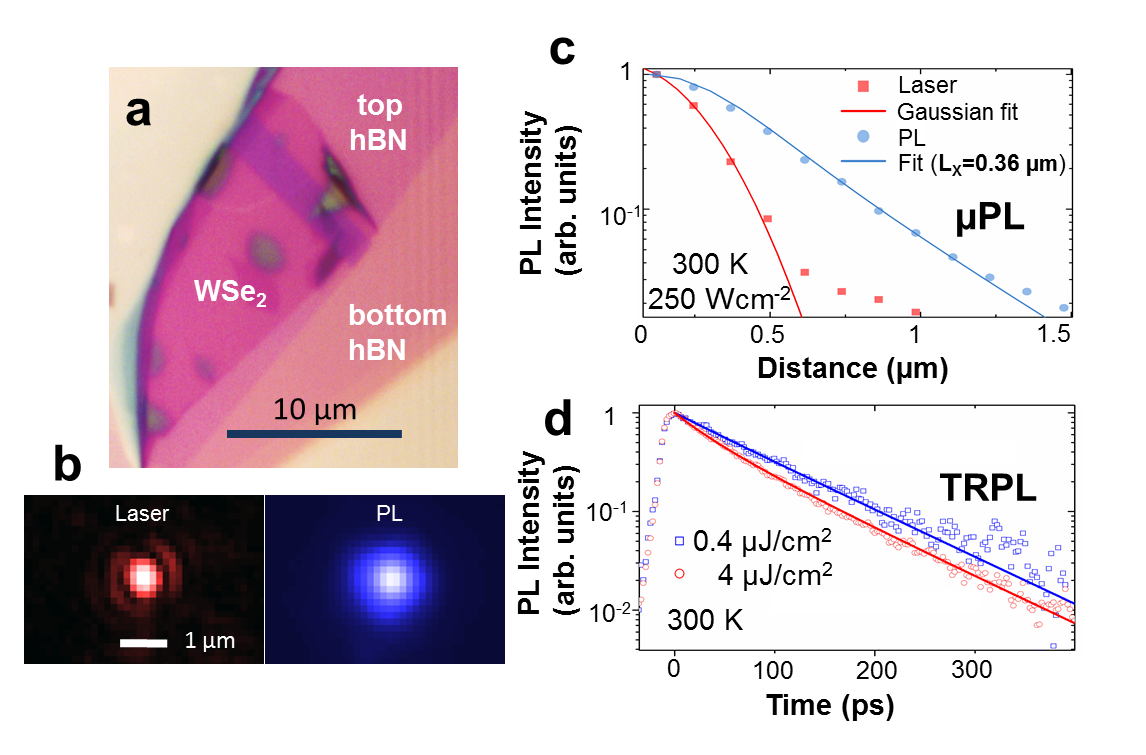}
\caption{\label{fig:fig1} (a) Optical microscope image of an WSe$_2$ flake encapsulated in hBN. (b) Image of the diffraction-limited spot used to perform µPL measurements (left), and resulting image of the PL at 300 K (right) (c)   Normalized photoluminescence intensity vs. distance to the excitation spot, revealing exciton diffusion at 300 K. The laser profile is also shown. (d) Time-resolved PL intensity at 300 K for two laser pulse energies ; the full lines are fits using equation (3) ; we find $\tau_X = 90 \pm 10$ ps and $\gamma N(0)= 7.7 \times 10^{-4}$ ps$^{-1}$ and $7.6 \times 10^{-3}$ ps$^{-1}$ for the two excitation powers. 
}
\end{figure}

\indent \textit{Samples and Experimental Set-up.---}
Van der Waals heterostructures such as the one shown in Panel (a) of Fig.\ref{fig:fig1} were fabricated by mechanical exfoliation of bulk WSe$_2$ crystals (2D semiconductors). A first layer of hBN is mechanically exfoliated onto a freshly cleaved SiO$_2$ (90 nm)/Si substrate using a viscoelastic stamp \cite{Gomez:2014a}. The deposition of the subsequent WSe$_2$ ML and the second hBN capping layer is obtained by repeating this procedure. After deposition of each layer, the sample is annealed at 150 $\degree$ C during 10 minutes.  \\

A micro-PL ($\mu$PL) set-up \cite{Favorskiy:2010a} is used to spatially monitor the exciton concentration in the temperature range $T=10-300$ K. MLs were excited with  a continuous wave He-Ne laser ($633$ nm) tightly focused onto a diffraction-limited spot in the sample plane. The resulting PL is imaged by a cooled Si-CCD camera after the laser is properly filtered. For TRPL measurements, the flakes were excited by a pulsed Ti:Sa laser (pulse width 1.5 ps, 80 Mhz repetition frequency, photon energy 1.784 eV, laser diameter of $\approx 1 \; \mu$m). The PL was dispersed by a spectrometer and detected with an Hamamatsu Streak camera \cite{Lagarde:2014a}.\\

\indent \textit{Results and Discussion.---}
Panel (b) of Fig.\ref{fig:fig1} shows the principle of the $\mu$PL experiment, which provides a contact-free optical approach to the measurement of the exciton diffusion length. A 633 nm, He-Ne laser is tightly focused onto the ML with a high numerical aperture (NA) objective. The resulting diffraction-limited spot can be fitted by a gaussian profile of the form $e^{-r^2/\omega^2}$, where $\omega =0.25 \;\mu m$ ($\omega =0.5 \;\mu m$) for a microscope objective of NA=0.95 at 300 K (resp. NA=0.75 at 10 K), as shown in Fig.\ref{fig:fig1}(c).  Since the resulting PL is spatially isotropic, a one-dimensional profile is obtained by averaging the PL intensity on the different radial directions. The result is shown in Fig.\ref{fig:fig1}(c) for an excitation power density of $250 \; W/\mbox{cm}^2$, where it can be seen that the excitonic luminescence comes from an spatial region which extends significantly beyond the excitation spot. For such power density, we have checked that we are in the linear regime of excitation where exciton-exciton annihilation effects can be neglected 
\cite{Amani:2015a, Robert:2016a}. In steady state conditions, the exciton concentration $n$ satisfies the simple diffusion equation:

\begin{equation}
\frac{P \alpha}{2\pi h\nu \omega^2} e^{-r^2/\omega^2} = \frac{n(r)}{\tau_X} - D_{X} \Delta n (r) 
\label{eq1}
\end{equation}

\noindent
where $P$ is the incident power, $\alpha$ the absorption coefficient at the photon energy $h\nu$, $\tau_X$ and $D_X$ are the exciton lifetime and diffusion coefficient, respectively. In a 2D crystal, Eq.(\ref{eq1}) can be solved analytically, and the solution is given by the convolution between the laser's gaussian profile and the modified Bessel function of the second kind $K_0$:

\begin{equation}
 n(r) \propto  \int_{-\infty}^{\infty} K_0(r/L_X) e^{-(r-r')^2/\omega^2} dr'   
 \label{eq2}
\end{equation}
\noindent
where $L_X=\sqrt{D_X \tau_X}$ is the exciton diffusion length. For a laser radius of $\omega=0.25\;\mu$m, the resulting PL profile can be fitted by using Eq.(\ref{eq2})  where the diffusion length is the only adjustable parameter. The fit gives $L_X= 0.36 \pm 0.02 \; \mu$m and it does not significantly vary with excitation power up to $50\; \mu$W. In order to determine all the relevant transport parameters, time-resolved photoluminescence was performed in order to determine $\tau_X$. Note that in TRPL experiments the initial photo-generated excitons density is usually too large to neglect the effects of exciton-exciton annihilation.   Figure \ref{fig:fig1} (d) shows the PL decay for two laser pulse energies  of 0.4 and 4 $ \mu J/\mbox{cm}^{2}$. Increasing the excitation energy by a factor 10 results in a decrease of the exciton lifetime due to exciton-exciton annihilation effects \cite{Mouri:2014a,Kumar:2014b, Amani:2015a, Robert:2016a}. To take into account the latter effects, the TRPL curves were fitted with the solution of the following equation:
\begin{equation}
\frac{\partial n}{\partial t} = -\frac{n}{\tau_X}- \gamma n^2
\label{eq3}
\end{equation}
\noindent
where the fitting parameters are the exciton lifetime $\tau_X$, and the product $N(0)\gamma$; $N(0)$ is the initial exciton concentration and $\gamma$ is the exciton-exciton annihilation rate. Considering that the experiments in Fig.\ref{fig:fig1}(d) are performed with a ratio of factor 10 between $N(0)$, the fit gives an exciton lifetime of $\tau_X = 90 \pm 10$ ps.  \\

\begin{figure*}
\includegraphics[width=0.8\textwidth]{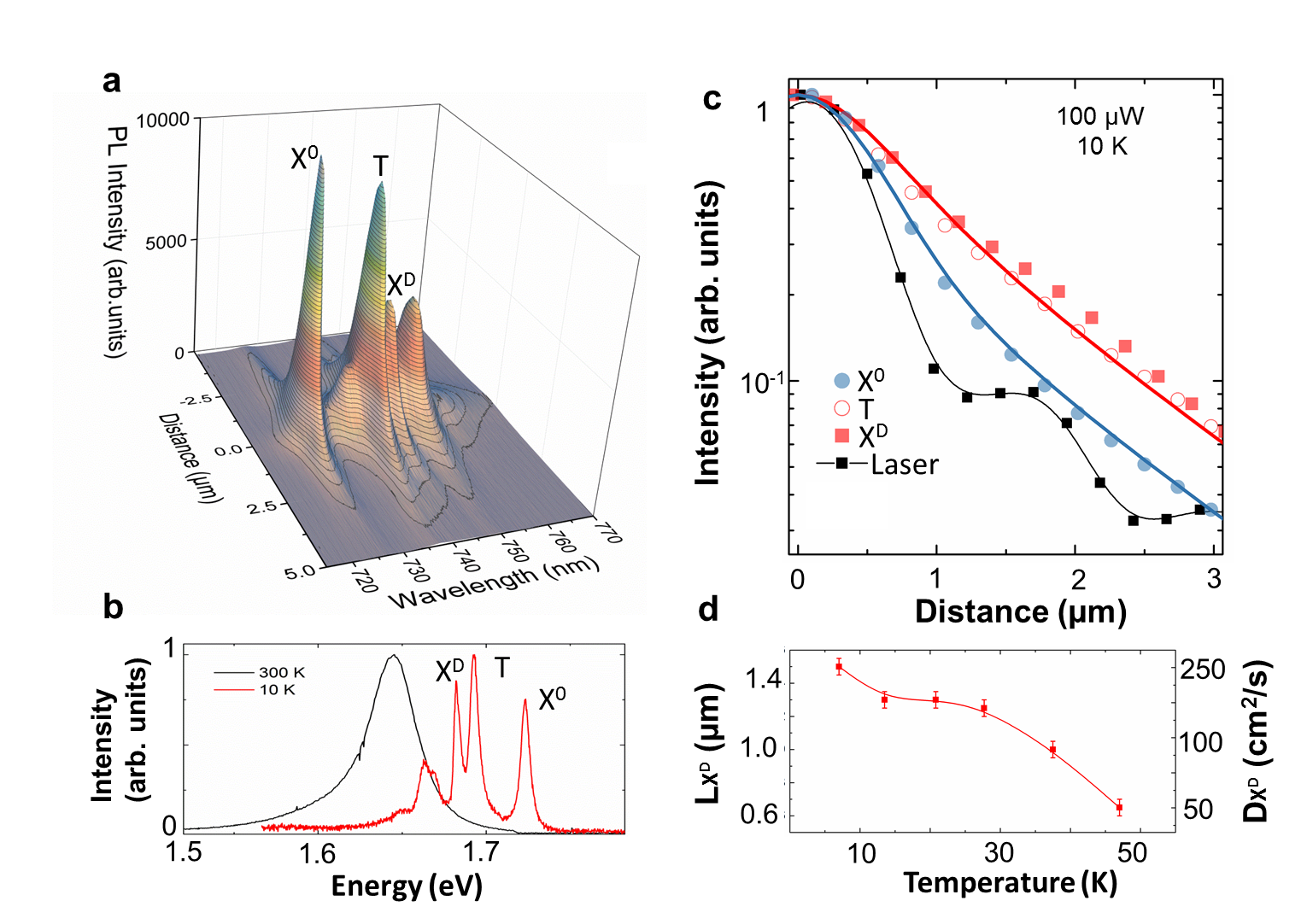}
\caption{\label{fig:fig2}  T=10~K. (a) 3D representation of the PL intensity as a function of the distance from the excitation spot centre and emission wavelength. (b) Spatially-integrated PL spectra at 300 K and 10 K. (c) PL intensity profiles for different excitonic complexes as a function of the distance to the excitation spot. Also shown is the laser profile. (d) Dark exciton diffusion length as a function of the lattice temperature. 
}
\end{figure*}

\noindent
By combining the $\mu$PL and TRPL results, we obtain an exciton diffusion coefficient of $D_X=14.5 \pm 2 \;\mbox{cm}^2$/s at 300 K, which is almost an order of magnitude larger than previously reported diffusion coefficients on WSe$_2$ monolayer flakes exfoliated onto SiO$_2$ \cite{Mouri:2014a,Yuan:2017a}, probably due to a high suppression of interfacial Coulomb scattering by charge impurities and remote scattering from phonons of the substrate after capping with hBN. The obtained value is comparable to a previously reported one measured by transient differential reflectivity in ML WSe$_2$ \cite{Cui:2014a} and to the one measured for excitons in bulk WS$_2$ \cite{Yuan:2017a}. By using the Einstein relation, the exciton diffusion coefficient can be written as:

\begin{equation}
D_{X} = \frac{\tau_c k_B T}{M_X}
\label{eq4}
\end{equation}

\noindent
where $k_B$ is Boltzmann's constant, $T$ the exciton temperature, $M_X$ is the exciton translational mass \cite{Konabe:2014a}, and $\tau_c$ the time interval between collisions that modify the total momentum of the excitonic ensemble. On the basis of recent magneto-absorption experiments performed on encapsulated monolayers \cite{Stier:2017a} and ARPES measurements \cite{Yuan:2016a}, we can estimate  $M_X \sim 0.8   \;m_0$. 
 By using the measured diffusion coefficient and $T=292$ K, one obtains $\tau_c \sim 260$ fs and therefore an effective exciton mobility of $\mu_X=q \tau_c/M_X \sim 576 \; \mbox{cm}^2/\mbox{V}/\mbox{s}$ where $q$ is the absolute value of the electron charge. The value obtained for $\tau_c$ is consistent with a very rough estimate of the dephasing time $T_2 \sim 90 $ fs that would correspond to an homogeneous linewidth of $\sim$ 15 meV at 300 K.  Recent four wave mixing experiments have revealed an homogeneous linewidth of 11 meV in WS$_2$ at 200 K\cite{Jakubczyk:2017a} and of 8 meV at 120 K in MoSe$_2$ \cite{Jakubczyk:2016a} deposited onto SiO$_2$ substrates, while very recent reflectivity measurements provides an estimation of $\sim$ 20 meV homogeneous linewidth in hBN capped MoSe$_2$ at 300 K. \\

\noindent
Now we switch to the investigation of exciton transport at cryogenic temperatures. Here, the excellent optical quality of hBN encapsulated monolayers \cite{Cadiz:2017a} allows to distinguish several  peaks corresponding to different excitonic complexes at the lattice temperature $T_L=10$ K. A typical spatially-integrated spectrum is shown in Fig.\ref{fig:fig2}(b). In order to discriminate the transport properties of the different peaks, an image of the PL is formed at the entrance slit of a grating spectrometer. The slit selects a vertical line on the image plane of the PL that passes trough the center of the PL spot.   The signal is then dispersed by a 600 gr/mm grating and collected by the CCD camera. The entrance slit's width is chosen to be $20\;\mu$m, i.e. equal to the CCD pixel size. \\

\noindent
Figure \ref{fig:fig2}(c) shows the spatial profile of the bright neutral exciton $X^0$,  the dark neutral exciton (z-polarized) $X^D$  \cite{Wang:2017c,Robert:2017a}and the trion (charged exciton)  $T$  \cite{Courtade:2017a}. The different transport dynamics are clearly resolved and we extract a diffusion length for the dark neutral exciton and for the trion of $L_{X^D}\approx L_{T}= 1.5 \pm 0.02 \;\mu$m. For the bright neutral exciton, the spatial decay cannot be accounted for with just one diffusion length. By allowing two lengthscales for $X^0$, one extracts a short decay ($L_X^1= 0.2 \pm 0.05 \;\mu$m) followed by a longer one ($L_X^2=1.5 \pm 0.04 \;\mu$m)\cite{Cadiz:2014a}. TRPL measurements have shown that the dark exciton in encapsulated WSe$_2$ at low temperatures has a lifetime of $\tau^{D}= 110 \pm 10$ ps \cite{Robert:2017a}, corresponding therefore to a diffusion constant of $D_{X^D}= L_{X^D}^2/\tau^D = 205 \pm 15 \;\mbox{cm}^2/s$. The use of the Einstein relation may not be pertinent since thermal equilibrium is not guaranteed here and the role of localization effects at cryogenic temperatures is an open question yet to be clarified.

 When the sample temperature is increased, there is a monotonic drop of the dark exciton diffusion length, as shown in Fig. \ref{fig:fig2}(d). 
 TRPL measurements have shown a weak variation of the dark lifetime up to 50 K \cite{Robert:2017a}, so that this decrease reflects mostly a significant reduction of the diffusion coefficient with temperature. Between 15 and 50 K, there is therefore a 4-fold decrease of the diffusion coefficient.   \\


\indent \textit{}
In conclusion, we have determined the transport parameters for excitons both at room and cryogenic temperatures. For the first time, diffusion of different excitonic species at low temperatures is resolved thanks to the excellent optical quality of TMDC MLs embedded in hBN. 
Future measurements in gated samples would provide  a powerful way to investigate drift and diffusion of different excitonic species (negatively and positively charged trions) and reveal the role of a background charge density on the effective mobility of the carriers. Polarization-resolved $\mu$PL can be also a significant tool to study the transport properties of the valley degree of freedom. Novel transport phenomena in TMDCs such as spin/valley-Hall effect with charged excitons, spin/valley-Coulomb drag \cite{Weber:2005a} and valley-dependent diffusion in degenerately doped TMDC MLs  \cite{Cadiz:2013a}, or in dense interacting and polarized neutral exciton gas, may be demonstrated with such experiments.\\

\indent \textit{Acknowledgements.---} 
We thank ANR MoS2ValleyControl and ERC Grant No. 306719 for financial support. X.M. also acknowledges the Institut Universitaire de France. C.R.  acknowledges ANR Vallex and Labex NEXT. M.M. aknowledges ITN Spin-NANO Marie Sklodowska-Curie Grant agreement No. 676108. K.W. and T.T. acknowledge support from the Elemental Strategy Initiative conducted by the MEXT, Japan and JSPS KAKENHI Grant No. JP15K21722.


\end{document}